\documentstyle[aps,pre,epsf]{revtex}

\begin{document}
\draft
\title{Jarzynski equality for the transitions between nonequilibrium 
steady states}
\author{Takahiro Hatano}
\address{Department of Pure and Applied Sciences, 
University of Tokyo, Komaba, Tokyo 153-8902, Japan}
\date{\today}
\maketitle

\begin{abstract}
Jarzynski equality [Phys. Rev. E {\bf 56}, 5018 (1997)], 
which has been considered to be valid for the transitions 
between equilibrium states, is found to be applicable to 
the transitions between nonequilibrium stationary states 
satisfying certain conditions.
Also numerical results confirm its validity.
Its relevance for nonequilibrium thermodynamics of 
the operational formalism is discussed.
\end{abstract}


%%%%%%%%%%%% INTRODUCTION %%%%%%%%%%%%%%

The framework of nonequilibrium thermodynamics has been 
sought by many authors \cite{jou} in order to treat 
various nonequilibrium systems such as chemical reactions, 
transport processes in solids, moleculer motors, etc.
Insofar, all these attempts seem to be based on 
the fluid-dynamical approaches, which mostly has the assumption 
of local equilibrium at its starting point.
Recently, Oono and Paniconi \cite{oono} present 
a different type of nonequilibrium thermodynamics whose 
framework corresponds to equilibrium thermodynamics.
The unique feature of their work lies in the fact that it is 
a set of laws concerning the operation from outside, as well as 
equilibrium thermodynamics.
We refer their theory as the operational formalism.
This formalism is so important that the concept of entropy 
in equilibrium thermodynamics is introduced concerning with 
the adiabatic operation \cite{lieb}.
The relation between dynamical entropy and thermodynamic entropy 
is also discussed from this viewpoint \cite{sasa}.
Hence it is interesting to construct nonequilibrium thermodynamics 
from the operational point of view, apart from the existing 
fluid-dynamical approach.

Operation by outside can cause energy exchange 
between the system and the external operator.
In equilibrium thermodynamics, there is a principle of 
the minimum work for the system in the isothermal environment; 
\begin{equation}
\label{minwork}
\Delta F\le\langle W\rangle , 
\end{equation}
where $\Delta F$ denotes the free energy difference between 
the initial state and the final state of the system, 
and $W$ denotes the work done by the external operator.
The average of a physical quantity $f$ is written 
as $\langle f\rangle$, as usual.
Note that the sign of $W$ is positive when the work is performed 
on the system.
The equality holds when and only when the process is reversible.
Jarzynski recently proposed the intriguing equality 
for the finite time transition between the equilibrium 
states \cite{jarzynski}; 
\begin{equation}
\label{jequality}
\exp(-\beta\Delta F)=\langle\exp(-\beta W)\rangle ,
\end{equation}
where $\beta$ denotes the inverse temperature.
Crooks \cite{crooks} gives another intriguing derivation of 
Eq. (\ref{jequality}) using the fluctuation theorem \cite{gallavotti}.
This equality is confirmed to be valid in the finite time transition 
between equilibrium states.
In this Rapid Communication, however, we show that Eq. (\ref{jequality}) 
is indeed applicable to the finite time transition between 
nonequilibrium steady states which satisfy certain conditions.
The derivation is given below by roughly following Ref. \cite{jarzynski}.

%%%%%%%%%%% DERIVATION %%%%%%%%%%%%%%%%

Consider the system with the following Hamiltonian; 
\begin{equation}
\label{hamiltonian}
{\cal H}= H_0(p)+H(x; \alpha)-xF(t), 
\end{equation}
where $\alpha$ is a parameter, and $F(t)$ denotes 
the purturbative driving force which may be 
responsible for the nonequilibrium situation, 
and $H_0(p)$ is independent of time.
The external agent manipulates the system 
by varying the parameter $\alpha$.
The system may be in contact with a heat bath or 
several heat baths of different temperatures.
In any cases, we describe the dynamics of the system 
by the stochastic process in the phase space spanned by $x$ and $p$.
We introduce the probability distribution function $f(\Gamma,t)$ and 
the transition probability $P(\Gamma,t|\Gamma',t')$, 
where $\Gamma$ denotes both $x$ and $p$, to get 
\begin{equation}
\label{phasespace1}
f(\Gamma,t)=\int d\Gamma'P(\Gamma,t|\Gamma',t')f(\Gamma',t').
\end{equation}
This leads to 
\begin{equation}
\label{phasespace2}
\frac{\partial f(\Gamma,t)}{\partial t}=
\int dx'R(\Gamma|\Gamma';t)f(\Gamma',t), 
\end{equation}
where 
\begin{equation}
\label{defineR}
R(\Gamma|\Gamma';t)=\lim_{\Delta t\rightarrow +0}
\frac{P(\Gamma,t+\Delta t|\Gamma',t)-P(\Gamma,t|\Gamma',t)}{\Delta t}.
\end{equation}
The dynamics of our nonequilibrium system is described by 
Eq. (\ref{phasespace2}) together with the initial condition.
Then we make an important assumption that the steady state of 
our system is characterized by the following distribution function;
\begin{equation}
\label{steady1}
f_{steady}(\Gamma;\alpha)\propto \Phi(x,p)\exp[-{\bar\beta}H(x;\alpha)], 
\end{equation}
where $\Phi(x,p)$ is an arbitrary function of $x$ and $p$, 
and ${\bar\beta}$ is a parameter that should be regarded as 
the effective inverse temperature.
In other words, we confine the theory to the systems 
whose stationary distribution functions are represented by 
Eq. (\ref{steady1}).
By the definition of the stationary state, 
Eq. (\ref{phasespace2}) leads to; 
\begin{equation}
\label{steady2}
\frac{\partial f_{steady}}{\partial t}=
\int d\Gamma'R(\Gamma|\Gamma';t)\Phi(x',p')
\exp[-{\bar\beta}H(x';\alpha)]=0.
\end{equation}

Our goal is to obtain the steady state version of 
Eq. (\ref{jequality});
\begin{equation}
\label{goal}
\langle\exp(-{\bar\beta}W)\rangle=\exp(-{\bar\beta}\Delta F), 
\end{equation}
while the meaning of $\Delta F$ is unclear at this point.
Note that ${\bar\beta}$ is identical to the one appearing 
in the distribution function Eq. (\ref{steady1}).
Adopting the path-integral expression, we write; 
\begin{equation}
\label{symbolic}
\langle\exp(-{\bar\beta}W)\rangle =
\int{\cal D}\Gamma (t)\exp (-{\bar\beta}W){\cal P}[\Gamma (t)], 
\end{equation}
where ${\cal P}[\Gamma (t)]$ is a probability distribution 
functional of the path $\Gamma (t)$ in the phase space.
The work done to the system is defined as \cite{sekimoto1} 
\begin{equation}
\label{work}
W=\int dt {\dot\alpha}\frac{\partial H(x; \alpha)}{\partial\alpha}.
\end{equation}
We manipulate the system by changing the value of $\alpha$ from 
$\alpha(0)$ to $\alpha({\cal T})$.

Then we discretize time duration of the operation $[0, {\cal T}]$ 
as $(t_0, t_1, \cdots, t_N)$, and write $\Gamma (t_i)$ as $\Gamma_i$ 
and ${\cal T}/N$ as $\Delta t$, respectively.
As a result of the discretization, the distribution functional 
${\cal P}[\Gamma(t)]$ is represented in terms of transition 
probability as follows; 
\begin{equation}
\label{functional}
{\cal P}[\Gamma(t)]=P_N(\Gamma_N|\Gamma_{N-1})\cdots
P_1(\Gamma_1|\Gamma_0)f_0(\Gamma_0),
\end{equation}
where $f_0(\Gamma_0)$ denotes the initial probability 
distribution function.
Similarly, Eq. (\ref{work}) becomes 
\begin{equation}
\label{discretework}
W=\sum_{i=0}^{N-1}\delta H_{i+1}(x_i), 
\end{equation}
where 
\begin{equation}
\delta H_{i+1}(x_i)=H(x_i;\alpha_{i+1})-H(x_i;\alpha_{i}).
\end{equation}
Due to Eqs. (\ref{functional}) and (\ref{discretework}), 
Eq. (\ref{symbolic}) is rewritten as 
\begin{equation}
\label{pathint2}
\langle\exp(-{\bar\beta}W)\rangle 
=\big[\prod_{i=0}^N\int d\Gamma_i\big]P_{N}(\Gamma_N|\Gamma_{N-1}) 
e^{-{\bar\beta}\delta H_N(x_{N-1})}\cdots P_1(\Gamma_1|\Gamma_0)
e^{-{\bar\beta}\delta H_1(x_0)}f_0(\Gamma_0).
\end{equation}
The integrals on the right-side of Eq. (\ref{pathint2}) are 
represented by the following iteration. 
\begin{equation}
\label{gi}
g_{i+1}(\Gamma)=\int d\Gamma_iP_{i+1}(\Gamma|\Gamma_i)
e^{-{\bar\beta}\delta H_{i+1}(x_i)}g_i(\Gamma_i), 
\end{equation}
where 
\begin{eqnarray}
\label{g0}
g_0(\Gamma)&=&f_0(\Gamma), \\
\label{gN}
\langle\exp(-{\bar\beta}W)\rangle &=&\int g_N(\Gamma)d\Gamma .
\end{eqnarray}
By taking the first order terms of $\Delta t$, we have 
\begin{eqnarray}
\label{P}
P_{i+1}(\Gamma|\Gamma_i)&=&\delta(\Gamma-\Gamma_i)
+\Delta tR_i(\Gamma|\Gamma_i)\\
\label{e}
e^{-{\bar\beta}\delta H_{i+1}(x_i)} &=&
1-{\bar\beta}\delta H_{i+1}(x_i)
\end{eqnarray}
Substituting Eqs. (\ref{P}) and (\ref{e}) into the recursive relation 
Eq. (\ref{gi}), and taking the limit $\Delta t \rightarrow 0$, we get 
\begin{equation}
\label{gequation}
\frac{\partial g(\Gamma,t)}{\partial t}=
-{\bar\beta}{\dot\alpha}
\frac{\partial H(x;\alpha)}{\partial\alpha}g(\Gamma,t)
+\int d\Gamma'R(\Gamma|\Gamma';t)g(\Gamma',t).
\end{equation}
This equation gives 
\begin{equation}
\label{gprop}
g(\Gamma,t)\propto \Phi(x,p)\exp[-{\bar\beta}H(x;\alpha(t))], 
\end{equation}
noting that the second term of the right-hand side of 
Eq. (\ref{gequation}) vanishes by Eq. (\ref{steady2}).
Since Eq. (\ref{g0}) tells us that $g(\Gamma,0)$ is identical to 
the initial probability distribution function $f_0(\Gamma)$, 
the right-side of Eq. (\ref{gprop}) must have 
an appropriate normalization factor;
\begin{equation}
\label{g}
g(\Gamma,t)=\frac{\Phi(x,p)}{Z_0}\exp[-{\bar\beta}H(x;\alpha(t))],
\end{equation}
where 
\begin{equation}
Z_0=\int d\Gamma\Phi(x,p)\exp[-{\bar\beta}H(x;\alpha(0))].
\end{equation}
From Eq. (\ref{gN}), we finally obtain the desired quantity;
\begin{equation}
\label{final1}
\langle\exp(-{\bar\beta}W)\rangle=\int d\Gamma g(\Gamma,{\cal T})
=\frac{Z_{\cal T}}{Z_0}, 
\end{equation}
where 
\begin{equation}
Z_{\cal T}=\int d\Gamma\Phi(x,p)\exp[-{\bar\beta}H(x;\alpha({\cal T}))].
\end{equation}
Note that $Z_0$ and $Z_{\cal T}$ depend on only the value of 
$\alpha(0)$ and $\alpha({\cal T})$, respectively, so that 
they are the state variables.
Namely, the quantity $\langle\exp(-{\bar\beta}W)\rangle$ 
does not depend on the transition process but 
depend only on the initial and the final states.
Furthermore, if we define the free energy by 
\begin{equation}
\label{freeenergy}
F=-{\bar\beta}^{-1}\log Z, 
\end{equation}
Eq. (\ref{final1}) gives our goal Eq. (\ref{goal}), 
which is rewritten as; 
\begin{equation}
\label{final2}
\Delta F=-{\bar\beta}^{-1}\log[\langle\exp(-{\bar\beta}W)\rangle].
\end{equation}
This finishes the derivations of the steady state version 
of Jarzynski equality Eq. (\ref{goal}).
In this derivation, the restriction on the stationary distribution 
function, Eq. (\ref{steady1}), is imposed.
It is quite unknown at this point that the Jarzynski equality 
holds for the system whose stationary distribution function 
does not satisfy the condition.
Hereafter we check the validity of the results 
by numerical simulations on some concrete models.

%%%%%%%%%%% NUMERICAL SIMULATIONS %%%%%%%%

We consider two examples.
First we treat the uniform temperature system whose 
Hamiltonian is given by 
\begin{equation}
{\cal H}=\frac{p^2}{2}+\frac{k(t)}{2}x^2-xA\sin(\omega t).
\end{equation}
This is one of the simplest models of the nonequilibrium 
steady state driven by external force.
By changing $k(t)$, we can put work onto 
the nonequilibrium system.
Although the sinusoidal force performs work to the system, 
its contribution is a stationary dissipation which 
characterizes nonequilibrium states; 
following Ref. \cite{oono}, we call the work which stationarily 
dissipates "house-keeping work".
We do not count its contribution into the work.

We employ the Langevin dynamics as a model of the heat bath; 
\begin{equation}
\label{langevin1}
{\ddot x}+\gamma {\dot x}+k(t)x=A\sin(\omega t)+\xi(t), 
\end{equation}
where $\xi(t)$ is the Gaussian white noise satisfying 
\begin{equation}
\langle\xi(t)\rangle =0,\ \ \ \ 
\langle\xi(t)\xi(t')\rangle = 2\gamma\beta^{-1}\delta(t-t').
\end{equation}
The control parameter $k(t)$ is changed from $1/4$ to $1$ as 
\begin{equation}
\label{k}
k(t)=\frac{1}{4}\big(1+\frac{3t}{{\cal T}}\big), 
\end{equation}
where ${\cal T}$ denotes the time duration of the operation.

Let us discuss the statistical property of the stationary state.
The model Eq. (\ref{langevin1}) leads to a time-dependent Kramers 
equation which yields time-dependent distributions, if the forcing 
period $2\pi/\omega$ is longer than the relaxation time of the system.
However, since the operation process is much slower than 
the forcing period, we average out the sinusoidal motion 
to get the stationary distribution.
If the forcing period becomes comparable to the relaxation time, 
the response of the system cannot follow the forcing so that the 
distribution functions become Gibbsian in the high-frequency limit 
$1/\omega\rightarrow 0$.
Here we choose the parameter such that the relaxation time of 
the position $\tau_x\sim\gamma$ is longer than the forcing period, 
and the one of the momentum $\tau_p\sim\gamma^{-1}$ is shorter 
than the forcing period, i.e. $\gamma^{-1}\le 2\pi/\omega\le\gamma$.
We can expect that the distribution of the position $\chi(x)$ becomes 
Gibbsian and the one of the momentum $\pi(p)$ is non-Gibbsian 
in this parameter range.
The obtained $\chi(x)$ and $\pi(p)$ are shown in Fig. \ref{fig1}, 
where we can see that our expectation is realized;
\begin{equation}
\label{steadysim}
f(\Gamma;k)\propto\exp[-\beta k\frac{x^2}{2}]\pi_0(p).
\end{equation}
Note that this satisfies the condition of Eq. (\ref{steady1}).
Following Eq. (\ref{freeenergy}), $\Delta F$ is calculated as 
${\bar\beta}^{-1}\log 2$ for this process.
Then we check if Eq. (\ref{final2}) holds.
Since the distribution function is given by Eq. (\ref{steadysim}), 
${\bar\beta}$ in Eq. (\ref{final2}) corresponds to $\beta$.
The quantity to be forcused here, 
$-{\bar\beta}^{-1}\log\langle\exp[-{\bar\beta}W]\rangle$, 
are shown in Fig. \ref{fig2} together with $\langle W\rangle$.
As is clearly seen, while $\langle W\rangle$ changes its value 
depending on ${\cal T}$, 
$-{\bar\beta}^{-1}\log\langle\exp[-{\bar\beta}W]\rangle$ 
is an invariant with respect to the operation time ${\cal T}$, 
which has been proved to be a state variable.
As ${\cal T}$ gets larger, $\langle W\rangle$ seems to converge 
to a finite value, which is identical to 
$-{\bar\beta}^{-1}\log\langle\exp[-{\bar\beta}W]\rangle$; 
we can regard this quantity as $\Delta F$.
These facts clearly indicate the validity of our main result.

On the other hand, by tuning parameters, 
we can get different steady states 
whose distribution functions do not satisfy Eq. (\ref{steady1}).
In those cases, we found that our equlity no longer holds.
However the principle of the minimum work still seems to be valid.

Second, we consider the system in contact with two heat baths 
of different temperatures.
The model we treat here is two Brownian particles 
coupled via the linear interacting potential.
The Hamiltonian of the systems is;
\begin{equation}
{\cal H}=\frac{p_1^2}{2}+\frac{p_2^2}{2}+\frac{k}{2}(x-y)^2.
\end{equation}
And the dynamics is written as;
\begin{eqnarray}
\label{twobody1}
\ddot{x}+\gamma_1\dot{x}+k(t)(x-y)&=&\xi_1(t),\\
\label{twobody2}
\ddot{y}+\gamma_2\dot{y}+k(t)(y-x)&=&\xi_2(t).
\end{eqnarray}
Again $\xi_1(t)$ and $\xi_2(t)$ are the Gaussian white noise;
\begin{equation}
\langle\xi_i(t)\rangle =0, \ \ \ 
\langle\xi_i(t)\xi_j(t')\rangle=2\gamma_i\beta_i
\delta_{ij}\delta(t-t'), 
\end{equation}
where 
\begin{equation}
\delta_{ij}=\left\{
\begin{array}{@{\,}ll}
1, & \mbox{$ (i=j)$} \\
0. & \mbox{$ (i\neq j)$}
\end{array}
\right.
\end{equation}
This may be the simplest heat conduction system, 
which is of course in nonequilibrium.
This system has been intensively studied by Sekimoto \cite{sekimoto}, 
and was found to have the following distribution; 
\begin{equation}
\label{steadysim2}
f_{steady}(\Gamma;k)\propto\exp[-{\bar\beta}k\frac{(x-y)^2}{2}]
\exp[-{\bar\beta}\frac{p_1^2+p_2^2}{2}], 
\end{equation}
where 
\begin{equation}
\label{betabar}
{\bar\beta}=
\frac{\gamma_1+\gamma_2}{\gamma_1\beta_1+\gamma_2\beta_2}
\beta_1\beta_2.
\end{equation}
The steady state of the system hence satisfies the condition 
of Eq. (\ref{steady1}).
We again control the parameter $k(t)$ as given in 
Eq. (\ref{k}) and check if Jarzynski equality holds.
With the knowledge on the distribution function, 
$\Delta F$ is calculated to be ${\bar{\beta}}^{-1}\log 2$ again.
The numerical result using ${\bar\beta}$ of Eq. (\ref{betabar}) 
is shown in Fig. \ref{fig2}.
It is clear that Jarzynski equality is also valid 
in this heat conducting systems.

%%%%%%%%% SUMMARY AND DISCUSSION %%%%%%%%%%%

In this Rapid Communication, we derive the steady state 
version of Jarzynski equality and reconfirm its validity 
by numerical simulations.
The condition in which the equality holds is that 
the statinary distribution function is given by Eq. (\ref{steady1}).
Note that the principle of the minimum work immediately follows 
Eq. (\ref{final2}).
Namely, the principle of minimum work is also valid for the transition 
between nonequilibrium steady states.

However, the equality has clear limitation on its application.
The condition Eq. (\ref{steady1}) is rather crucial 
that it cannot describe the steady state of the system 
where the temperature depends on the position; 
e.g. the Brownian particle in the nonuniform temperature 
environment \cite{miki}.
It is unclear to what extent Eq. (\ref{steady1}) is satisfied 
in various nonequilibrium systems.

Another open question is the definition of the free energy 
in nonequilibrium systems.
As we have seen in the numerical simulations above, 
Eq. (\ref{freeenergy}) seems to be valid in the system 
satisfying Eq. (\ref{steady1}) regardless of the statistical 
property of the momentum space.
However, in more general systems, 
the definition of the nonequilibrium free energy as well as 
entropy is still unclear, although it may appear as 
the minimum work as is stated in Eq. (\ref{minwork}).
The broader application and the further development of 
the framework stated in Ref. \cite{oono} must be fruitful 
for nonequilibrium thermodynamics and should be the main focus 
of the future problem.

The author thanks S. Sasa for helpful discussions 
and his critical reading of the manuscript.
Discussions with S. Takesue, H. Nakazato, and Y. Yamanaka 
are also gratefully acknowledged.

\begin{figure}
\label{fig1}
%\begin{center}
%\epsfile{file=fig1.eps}
\caption{ 
The steady state distribution functions in the configuration space 
$\chi(x)$ (diamonds) and in the momentum space $\pi(p)$ (solid line).
Dashed line represents the Gaussian distribution corresponding 
to $\exp[-\beta kx^2/2]$.
Parameters are set as 
$k=1$, $\beta =1$, $\gamma =2$, $A=2$, and $\omega =3$.}
%\end{center}
\end{figure}

\begin{figure}
\label{fig2}
%\begin{center}
%\epsfile{file=fig2.eps}
\caption{
Averaged work and the free energy of the two examples in the text.
Plots on the down side are for the sine forcing system.
Blanked diamonds and circles denote $\langle W\rangle$ and 
$-{\bar\beta}^{-1}\log\langle\exp[-{\bar\beta}W]\rangle$ 
, respectively.
All the parameters are the same as Fig. 1.
Plots on the up side are for the heat conducting system.
Filled diamonds and circles denote $\langle W\rangle$ and 
$-{\bar\beta}^{-1}\log\langle\exp[-{\bar\beta}W]\rangle$ 
, respectively.
We set the parameters to be $\gamma_1=\gamma_2=1$, 
$\beta_1=0.5$, and $\beta_2=1$.
Dashed lines denote the free energy difference calculated from 
the stationary distribution functions.}
%\end{center}
\end{figure}

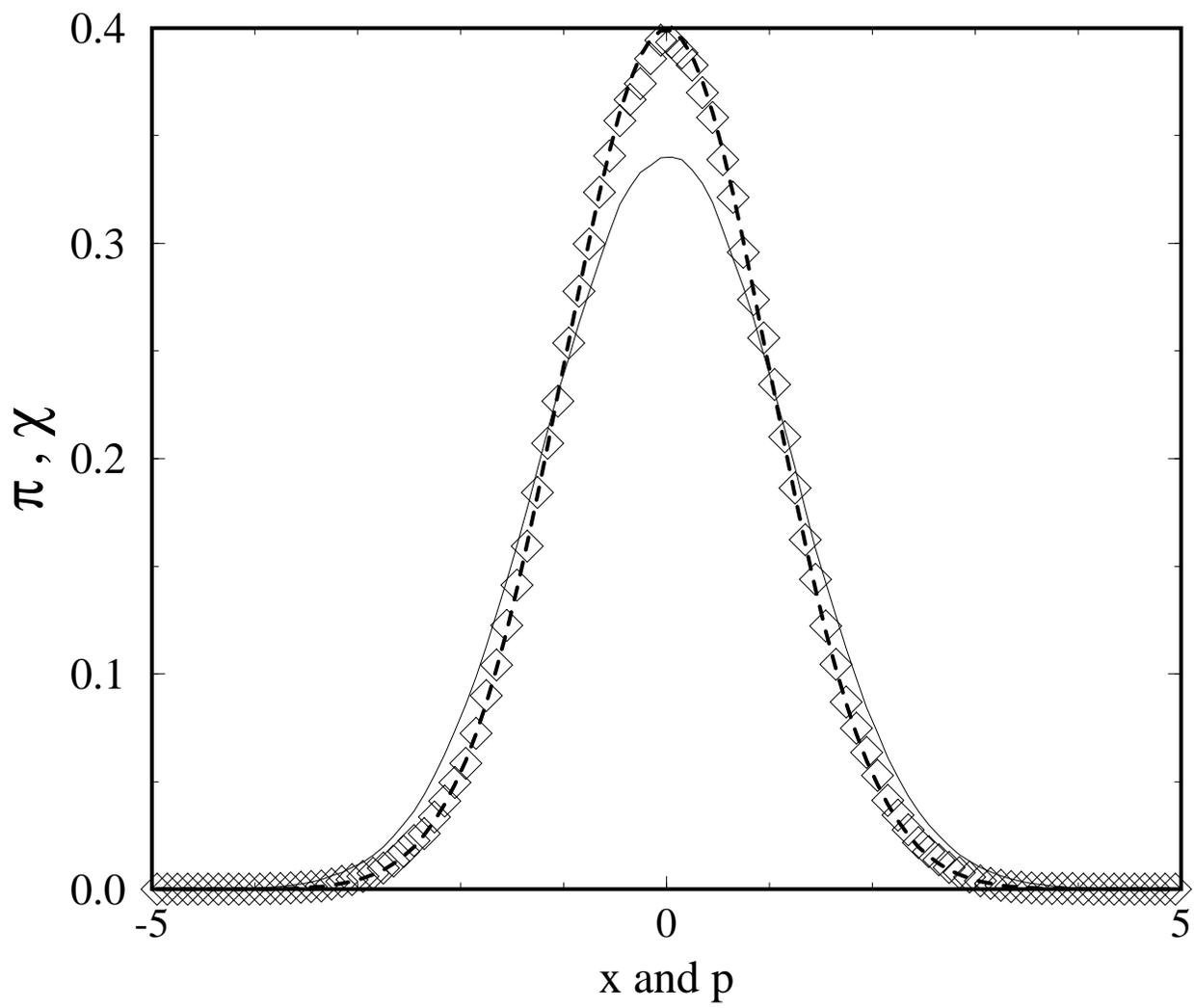
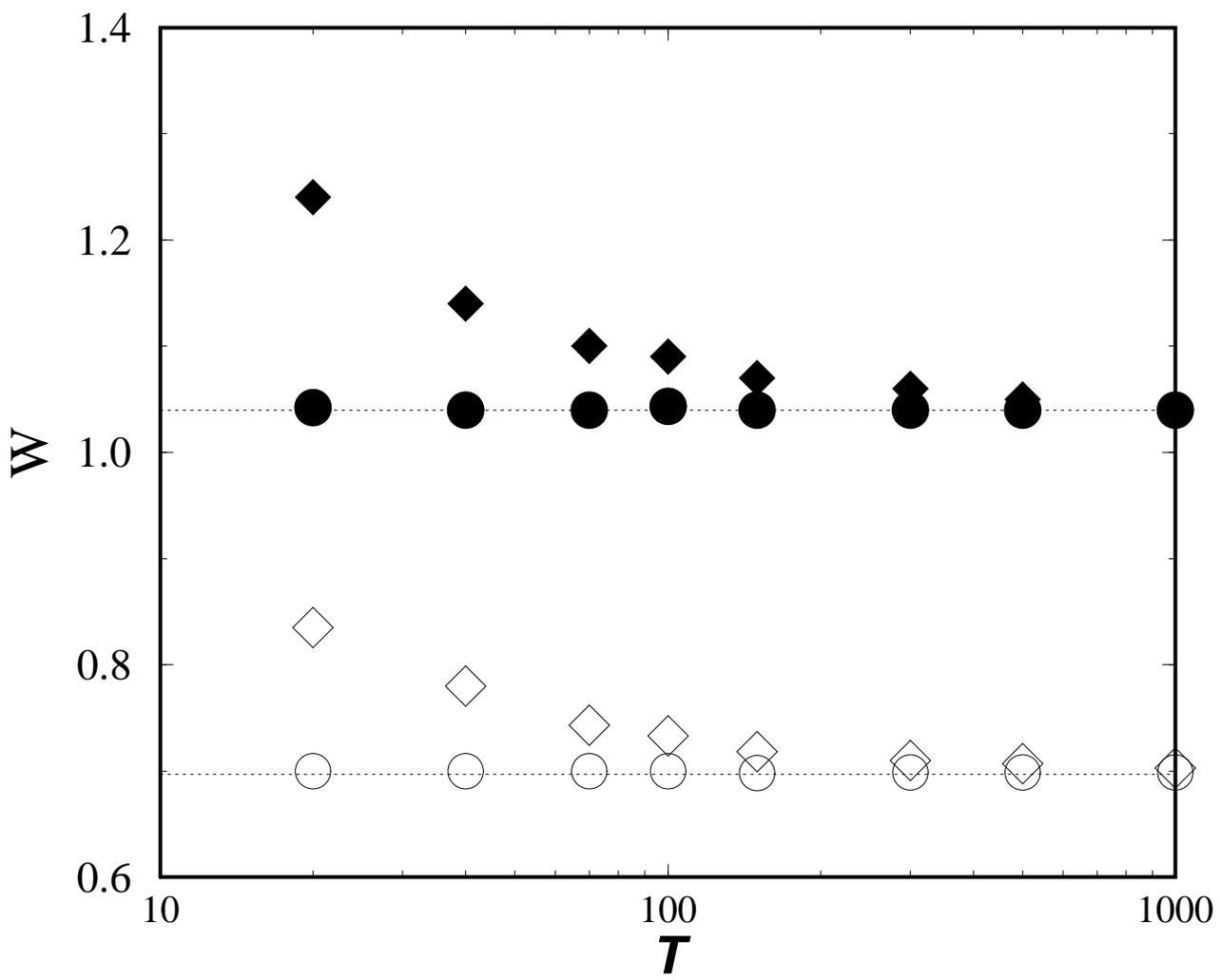
\end{document}